# Breaking the spatial resolution barrier via iterative sound-light interaction in deep tissue microscopy

Ke Si[§], Reto Fiolka[§] and Meng Cui[*]

*Howard Hughes Medical Institute, Janelia Farm Research Campus,
19700 Helix Drive, Ashburn, Virginia, 20147, USA*

[§] Equal contribution

[*]Correspondence and request for materials should be addressed to Meng Cui, 571-209-4136
cuim@janelia.hhmi.org

**Abstract**

Optical microscopy has so far been restricted to superficial layers, leaving many important biological questions unanswered. Random scattering causes the ballistic focus, which is conventionally used for image formation, to decay exponentially with depth. Optical imaging beyond the ballistic regime has been demonstrated by hybrid techniques that combine light with the deeper penetration capability of sound waves. Deep inside highly scattering media, the sound focus dimensions restrict the imaging resolutions. Here we show that by iteratively focusing light into an ultrasound focus via phase conjugation, we can fundamentally overcome this resolution barrier in deep tissues and at the same time increase the focus to background ratio. We demonstrate fluorescence microscopy beyond the ballistic regime of light with a threefold improved resolution and a fivefold increase in contrast. This development opens up practical high resolution fluorescence imaging in deep tissues.

Optical microscopy is an invaluable tool in the biological sciences[1-7] as it enables three-dimensional non-invasive *in vivo* imaging of the interior of cells and organisms with molecular specificity. Unfortunately optical methods are restricted to an imaging depth of a few scattering mean free path lengths[8-10], a severe limitation in many research fields[3, 11, 12]. Recently hybrid techniques[8, 13-18] that combine the deep penetration capability of sound waves and the molecular contrast of light waves have greatly exceeded the depth limitation of pure optical methods. However, at these extended depths the achievable spatial resolution is restricted by the dimensions of the sound focus. Here we present an approach to fundamentally break the resolution limit of hybrid imaging technologies in deep tissue. Through iterative ultrasound guided optical phase conjugation (OPC), we shrink the sound light interaction volume and obtain a drastically sharper optical focus. This technology paves the way for deep-tissue fluorescence microscopy for biological research and medical applications.

The shallow optical penetration depth has restricted many research fields: it has forced biologists to use transparent model organisms, monolayer cell cultures or histological sections of tissue, just to name a few compromises. Consequently a lot of effort was dedicated to push the depth range in optical imaging[10, 19-26] and recently substantial progress has been reported using hybrid approaches that combine light and sound[4, 8, 16]. Yet there is still a need for a technique that can take full advantage of the wealth of fluorescent labels and provide microscopic resolution at depths of 1mm in tissues or deeper. For this goal, we need the ability to focus light tightly beyond the ballistic regime at arbitrary locations.

Recently, light focusing deep inside tissues was achieved using ultrasound guided optical phase conjugation[13, 14] and fluorescence imaging was demonstrated with NIR[17] and visible[18] excitation. An ultrasound focus, which experiences much less scattering than light, is used as a source of frequency shifted light that can be recorded and time-reversed using OPC. Similar to other hybrid techniques, however, the resolving power at large depths is determined by the size of the ultrasound focus, resulting in modest spatial resolutions of 30-50 microns[17, 18]. Further the first demonstrations[17, 18] lacked sufficient contrast for practical biological imaging.

Here we demonstrate fluorescence microscopy beyond the ballistic regime with a lateral resolution of ~12 microns using iterative ultrasound guided digital OPC. We overcome the sound resolution limit by a factor of three and at the same time increase the focus to background ratio (FBR) fivefold. The principle behind our technique can be explained as follows: after traveling through highly scattering media, the incident light field at the ultrasound focus is completely randomized and unfocused. However, if the light was already pre-focused into the ultrasound focus using OPC, a much more confined sound-light interaction would occur.

Let us assume that the transverse profile of the sound modulation zone and hence the phase conjugation beam at the sound focus is defined as $M(y,z)$ and that we employ two digital optical phase conjugation (DOPC) systems[27], DOPC1 and DOPC2. DOPC1 first illuminates the sample and the sound modulated light is recorded by DOPC2, which is schematically shown in Fig. 1 **a**. DOPC2 then generates a phase conjugation beam that focuses back to the sound focus (Fig. 1 **b**). Different from the first illumination, the DOPC2 beam has a focused light distribution $M(y,z)$ at the sound focus. Therefore the emerging sound modulated light has a new spatial profile $M(y,z)^2$. If we let the two DOPC systems take turns to illuminate the sample and to record the sound modulated light, we can achieve a focus profile $M(y,z)^N$, where N is the iteration number (Fig. 1 **c-d** ).

If we assume a Gaussian profile for $M(y,z)$ and a strong optical focus (large FBR) for a single OPC operation, the transverse FWHM of the PSF decreases as $1/\sqrt{N}$. The FBR can be estimated by the number of independently controlled phase pixels $N_{pixel}$ of the SLM divided by the number of uncorrelated optical modes $N_{mode}$ present in the ultrasound focus[24, 28]. In a 2D approximation, the sound-light interaction area is reciprocally related to N and thus the FBR is expected to increase linearly with N. If the initial focus quality is low (FBR < 5), the dependence of FWHM and FBR on the number of iterations is more complicated. We use numerical simulations to investigate this regime, as described in the Supplementary discussion.

For fluorescence imaging, the ultrasound focus was raster scanned through the sample and at each position iterative DOPC was performed. The power of the fluorescence emission for each DOPC excitation was measured and the fluorescent background level was subtracted. The background signal was obtained by lateral translation of the DOPC phase pattern[17, 25, 28] (30 pixels in z and y), which makes the phase conjugation ineffective.

To demonstrate the resolution increase using iterative DOPC, we measured the three-dimensional PSF of our system. To this end, we embedded 6 micron diameter fluorescence beads in a slice of Agar (200 microns thick) and sandwiched the slice between two tissue phantoms (scattering coefficient: 7.63 /mm, g factor: 0.9013) of 2 mm in thickness. The details of the sample preparations are included in the supplementary discussion.

Figure 2 **a** shows the lateral PSF for DOPC iteration 1, 3, 5, 7, and 9. Figure 2 **b** shows the axial PSF for iteration 1, 5, and 9. To determine the full width half maximum (FWHM), Gaussian fitting through cross-sections of each PSF was applied (Fig. 2 **c-e**). For iteration 1, when DOPC is applied the first time, the mean FWHM of the PSF amounts to 35.7, 39.0 and 142 microns in the y, z and x (axial) direction, respectively (Fig. 2 **f-g**). After nine DOPC iterations, the FWHM was reduced to 11.2, 12.8, and 60.3 microns in the y, z, and x directions. The FBR is increased by a factor of ~5 over nine iterations and appears to grow almost linearly with N (Fig. 2 **h**). Besides the FBR, the total sound modulated light power increases as well (Fig. 2 **i**), however not linearly with N. We have simulated the iterative DOPC process (see Supplementary discussion) and the results are generally in good agreement with the experiments (Fig. 2 **f-i**). In addition, we also performed PSF measurements through 1.2 mm thick fixed rat brain tissue, as shown in Supplementary Fig. 1.

To demonstrate imaging of a complex fluorescent structure deep inside highly scattering media, we fabricated a c-shaped pattern consisting of fluorescent microspheres of 6 microns in diameter, completely embedded in the middle of a 4 mm thick tissue phantom (scattering coefficient: 7.63 /mm, g factor: 0.9013). In Fig. 3 **a**, a widefield microscopy image of the c-shaped fluorescence pattern is shown before it was embedded in the tissue phantom. Figure 3 **b** shows a widefield image taken through the tissue phantom. The shape information is completely lost due to the strong scattering. In Fig. 3 **c,** an image obtained with the first DOPC iteration is shown. The scanning step size was 6 microns and the raw data was re-sampled with linear interpolation. The structure can now be localized, but the shape of the object is not resolved. In Fig. 3 **d,** an image obtained after five DOPC iterations is shown. At this stage, the c-shape is already recognizable. After nine DOPC iterations the c-shape structure is clearly resolved owing to the increased lateral resolution (Fig. 3 **e**). For comparison, we re-sampled the widefield image in Fig. 3 **a** to the same pixel size as in Fig. 3 **c-e** and convolved it with a Gaussian-shaped PSF (FWHM: 12 microns). The resulting image is shown in Fig. 3 **f**. An additional imaging experiment using sparsely distributed fluorescent beads is shown in Supplementary Fig. 2.

In this study, we break the sound wave limited resolution barrier in the diffusive regime through iterative sound modulated DOPC. This technique improves the resolution in deep tissue fluorescence microscopy towards 10 microns while increasing the focus to background ratio by a factor of 5 at the same time. Better SLM performance such as lower pixel coupling and reduced phase jitter is expected to improve the current single iteration FBR by more than an order of magnitude, potentially yielding an exact N fold FBR gain through iterations. The increase in sound modulated power allowed us to shorten the acquisition time for the wavefront recording after the first couple of iterations. Moreover, this effect may enable us to focus light even deeper into tissue: by translating the sound focus in small steps between the iterations, the light focus can be gradually guided into deeper regions while maintaining a high sound modulated signal level.

In conclusion, our development is an important step towards practical deep-tissue fluorescence microscopy, providing sufficient resolution and contrast for many applications. Further improvement is expected with two photon fluorescence excitation, potentially leading to sub 10 micron spatial resolution and FBR > 200. We envision that our technique will find numerous applications in neuroscience, optogenetics, medical diagnostics, photodynamic therapy and other fields that require localized light radiation deep inside tissues.

## Methods

### Setup

Figure 1 **e** shows our experimental setup: two identical DOPC systems are used either to illuminate the sample with a phase conjugated beam or to record a wavefront emanating from the ultrasound focus within the sample. A Q-switched laser pumped Ti:sapphire oscillator, centered at 778 nm and with 20 ns pulse duration (Photonics Industries, NY), is split into two beams for the two DOPC systems. The two laser beams are used to illuminate the sample via DOPC1 and to serve as a reference beam to record a wavefront on DOPC2 or vice versa. In the beam path of DOPC2, the light is frequency shifted using an acousto-optical modulator such that a 10 Hz beating between the reference beam and the light emanating from the ultrasound guide star results when either DOPC system is used for wavefront recording. This beating is recorded by the camera of either DOPC system, allowing us to recover the wavefront using phase stepping interferometry. Since the laser has a finite coherence length (~ 1 cm), the path length has to be adjusted depending on which DOPC is used for wavefront recording to ensure proper interference. To this end, the optical path length for DOPC1 can be rapidly switched using beamsplitters and two fast mechanical shutters.

The sample is housed in a water chamber with three optical windows. Below the sample, an ultrasound transducer is mounted on a 3-axis motorized stage. Fluorescence emission is filtered by a bandpass filter and is imaged from the side of the sample chamber onto a camera. The camera is not used to record a spatially resolved widefield image but to measure the power of the fluorescence emission by summing all of its pixels. To form a fluorescence image, the ultrasound focus is raster scanned through the sample and at each position, iterative DOPC is applied. For each applied phase conjugation, the fluorescence emission is recorded with the camera. The timing and synchronization scheme was described in a previous publication[17].

## Competing financial interests

The authors declare no competing financial interests.

## Author contributions

The experiment was designed and implemented by M.C. Image data was acquired by M.C. and R.F. The fluorescence pattern was created by K.S. The scattering coefficient was measured by R.F. The numerical simulation was performed by K.S. All authors contributed to the data analysis and the preparation of the manuscript.

## Acknowledgements

We are grateful to Brenda Shields and Amy Hu for preparing the rat brain slices and Doug Murphy for lending us a filter set. We thank Mats Gustafsson for his insightful questions during the planning stage of this project. The research is supported by the Howard Hughes Medical Institute.

Figure 1

**a-d** Schematic illustration of the iterative focus improvement. **a** The initial incident light field (purple) propagates to the ultrasound focus (yellow circle). A simulated speckle pattern at the sound focus (location marked with the white arrows) is shown in the right inset. A small portion of the input light is frequency shifted (green). **b** In the first DOPC iteration, the green light field is time-reversed and is re-focused into ultrasound focus, resulting in a more confined sound light interaction (right inset). A portion of the light is frequency shifted (purple). **c** The purple light field is time reversed and is re-focused into the ultrasound zone, further shrinking the sound-light interaction zone. **d** After nine iterations, the time-reversed purple light field results in a much improved focus. **e** Experimental setup; PO, Pockels cell; I, Isolator; BS, beam splitter; AOM, acousto-optical modulator; ND, neutral density filter, BB, beam block; DL, delay line; BE, beam expander; P, polarizer; BP, bandpass filter; L1, f = 35 mm lens; L2, f = 50 mm lens; D, fluorescence detector. Scalebar: 10 microns.

Figure 2

**a** Lateral PSF measurement through 2 mm thick tissue phantoms ($\mu_s$ = 7.63 /mm, g factor = 0.9013) for iterations 1, 3, 5, 7, and 9. To normalize the peak intensity, the PSF data sets were multiplied by 6.5, 2, 1.5, and 1.5 for iteration 1, 3, 5, and 7, respectively. **b** Axial PSF measurements for iterations 1, 5, and 9. The PSF data for iteration 1 and 5 was multiplied by 6.5 and 1.5, respectively. **c-e** Gaussian fitting of the measured PSF. **f** Fitted transverse FWHM and simulation (mean values and standard deviation). **g** Fitted axial FWHM and simulation (mean values and standard deviation). **h** Measured focus to background ratio and simulation (mean values and standard deviation). **i** Measured ultrasound modulated light power and simulation (mean values and standard deviation). Scalebar: 10 microns. Colorbar in arbitrary units.

Figure 3

**a** Direct widefield image of the fluorescent structure without tissue phantoms. **b** Direct widefield image of the fluorescent structure surrounded by 2 mm thick tissue phantoms ($\mu_s$ = 7.63 /mm, g factor = 0.9013). **c** Image acquired with the first round of ultrasound pulse guided DOPC. **d** Image acquired with five iterations. **e** Image acquired with nine iterations. **f** 2D convolution of **a** with a 2D Gaussian function (FWHM: 12 microns). Scalebar: **a**, **b**: 100 microns, **c**: 20 microns. Colorbar in arbitrary units.

Figure 1

Figure 2

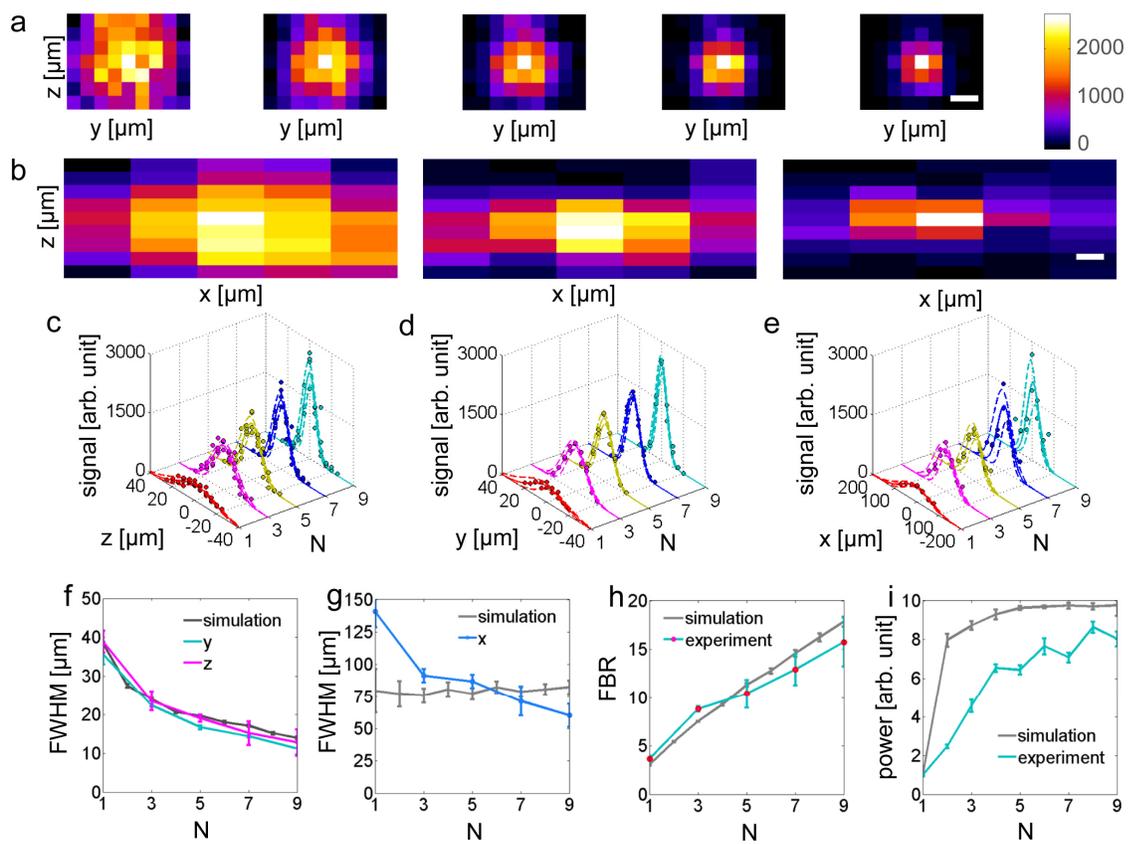

Figure 3

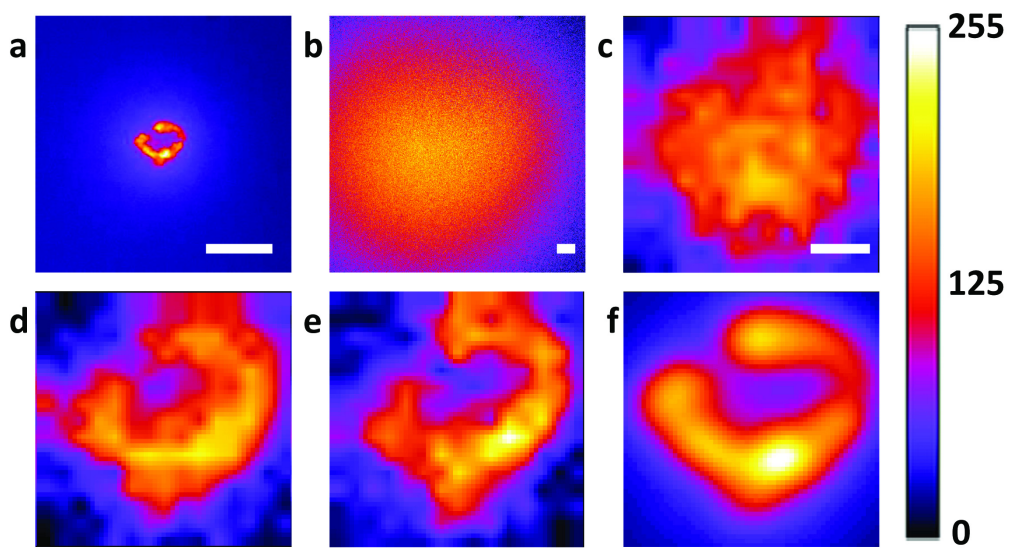

# Supplementary Discussion

**Numerical Simulation**

For a weak DOPC focus (FBR < 5), the background light generated by the OPC beam is not negligible and contributes to $N_{mode}$. We therefore suspect that in this case the FWHM and FBR may scale differently from $1/\sqrt{N}$ and N, respectively. In experiments, the FBR with a single OPC operation is < 5 due to the inter-pixel coupling and phase jitter of the SLM. To explore this regime, we use numerical simulations to compute the dependence of the FWHM, FBR, and the sound modulated light power on the iteration number N.

To mimic the random scattering media, we defined a 1.2x1.2x4 $mm^3$ volume filled with randomly distributed scatterers, whose volume density is 15% and average diameter is one micron. We assumed that the real part of the refractive index of the scatterers is randomly distributed from $n_{min}$ to $n_{max}$ and the imaginary part was zero. Therefore absorption is neglected, a reasonable assumption for NIR light in biological tissues. The value $n_{max}$ was controlled to yield the desired mean scattering path length. We divided the volume into 200 layers (phase masks). Light propagation between the phase masks was computed via the Fourier shift theorem. The associated Fourier transforms were computed with FFT in MATLAB. We simulated the light scattering process by multiplying the E field with the phase masks. Polarization effects were not incorporated in the simulation.

We assumed that the light-ultrasound interaction was a simple Doppler shift and we approximated the ultrasound focus as a two-dimensional Gaussian function in the middle of the scattering volume. The light reaching the sound focus was frequency shifted and its intensity was multiplied by the ultrasound intensity. After that, only the frequency shifted E field was further forward propagated. We assumed that the DOPC system was an ideal pixelated OPC mirror: the phase of the E field was resampled on a grid of 200x200 pixels (the number was chosen to match the FBR with a single OPC operation in the experiment). For phase conjugation, we reversed the sign and propagation direction of the E field. For every iteration, we assumed that the amplitude of the OPC beam was uniform and identical such that the incident light power on the sample always had the same value.

Using these "building blocks", we ran simulations of iterative ultrasound guided DOPC and obtained the E field distribution at any desired location. To compute the PSF we convolved the focused light intensity distribution with a 6 micron diameter bead, as used in the experimental PSF measurements. From the resulting data, we determined the FWHM and the FBR. We computed the ultrasound modulated light power from the summation of the frequency shifted light intensity on the OPC mirror. We repeated all the simulations using different randomly generated scattering media with the same average scattering coefficient and g factor.

Our simulation does not accurately predict the decay of the axial FWHM (see Fig. 2 **g**), likely due to the 2D approximation of the sound-light interaction. Since all the other simulated values are in good agreement with the experiments, we did not expand our simple model, but note its limitations in predicting the 3D light intensity distribution.

**Sample Preparation**

We prepared the tissue phantoms used for Fig. 2 and 3 by mixing one micron diameter polystyrene beads suspension (2.6% solid) with Agar at 80:920 volume ratio. The scattering coefficient was measured using a previously described method[1] and amounts to 7.63 /mm, (95% confidence bounds: 7.46-7.79 /mm). We calculated the scattering anisotropy factor using a Mie scattering calculator[2]. The value is 0.9013 with the assumption $n_{water}$ = 1.330 and $n_{polystyrene}$ = 1.579. To prepare the c-shaped pattern, we manually punched a hole of 50 microns in diameter and 60 microns in depth on the surface of a 2 mm thick tissue phantom using a glass micropipette. We injected 6 micron diameter fluorescence beads into the hole. Through careful rinsing, we gradually removed the beads in the center until a c-shaped pattern remained. Afterwards, we sealed the hole with a 0.2 mm thick tissue phantom, and recorded a widefield fluorescence image (Fig. 3 **a**) using an Olympus IX71 fluorescence microscope as a reference. Finally, we added one 1.8 mm thick tissue phantom on top such that the fluorescence pattern was completely embedded in the middle of the 4 mm thick tissue phantom.

To measure the PSF through rat brain tissue (Supplementary Fig. 1), we prepared a 1 mm thick Agar layer containing a sparse distribution of 6 micron fluorescence beads. The slice was sandwiched between two 1.2 mm thick fixed rat brain slices. The scattering coefficient of the rat brain tissue was determined in a previous study[1] and amounts to 12.78 /mm (95% confidence bounds: 11.6-13.96 /mm).

For the sample shown in Supplementary Fig. 2, we dried a drop of 6 micron diameter fluorescence beads suspension on the surface of a 1 mm thick clear Agar slice. We then sealed the fluorescent beads with an additional 0.2 mm thick Agar layer. We selected a suitable bead pattern and cut out a small cube (~1x1x1.2 mm$^3$) containing this pattern. The cube was re-embedded inside a clear 2 mm thick Agar slice, which was subsequently sandwiched between two 2 mm thick tissue phantoms. We prepared these two tissue phantoms by mixing 1.5 micron diameter polystyrene beads suspension (2.61% solid) with Agar at a 66:934 volume ratio. The calculated g factor is 0.93056 and the scattering coefficient was determined in a previous publication[1] to be 6.42 /mm (95% confidence bounds: 6.286-6.555 /mm) .

**Supplementary References**

1. Si, K., Fiolka, R. & Cui, M. Fluorescence imaging beyond the ballistic regime by ultrasound-pulse-guided digital phase conjugation. Nature Photonics Advance online publication (2012).

2. Prahl, S. Mie Scattering Calculator http://omlc.ogi.edu/calc/mie_calc.html.

**Supplementary Figure 1**

PSF measurements through fixed rat brain slices. **a** Lateral PSF measurements for iteration 1, 3, 5, 7, and 9. To normalize the peak intensity in all images, the PSF data were multiplied by 7.2, 2.2, 1.4, and 1.2 for iteration 1, 3, 5, and 7, respectively. **b-c** Gaussian fitting of the measured PSF. **d** Fitted transverse FWHM (mean values and standard deviation). Scalebar: 10 microns. Colorbar in arbitrary units.

**Supplementary Figure 2**

Fluorescence imaging of 6 micron diameter beads sandwiched between 2 mm thick tissue phantoms ($\mu_s$ = 6.42 /mm, g factor = 0.9306). **a** Direct widefield image of the fluorescent beads without tissue phantoms. **b-e** Image of the fluorescent beads through 2 mm thick tissue phantoms for iteration 1, 3, 5, and 9, acquired with iterative ultrasound pulse guided DOPC. **f** Widefield image **a** convolved with a 2D Gaussian function (FWHM: 12 microns). Scalebar: 10 microns. Colorbar in arbitrary units.

Supplementary Figure 1

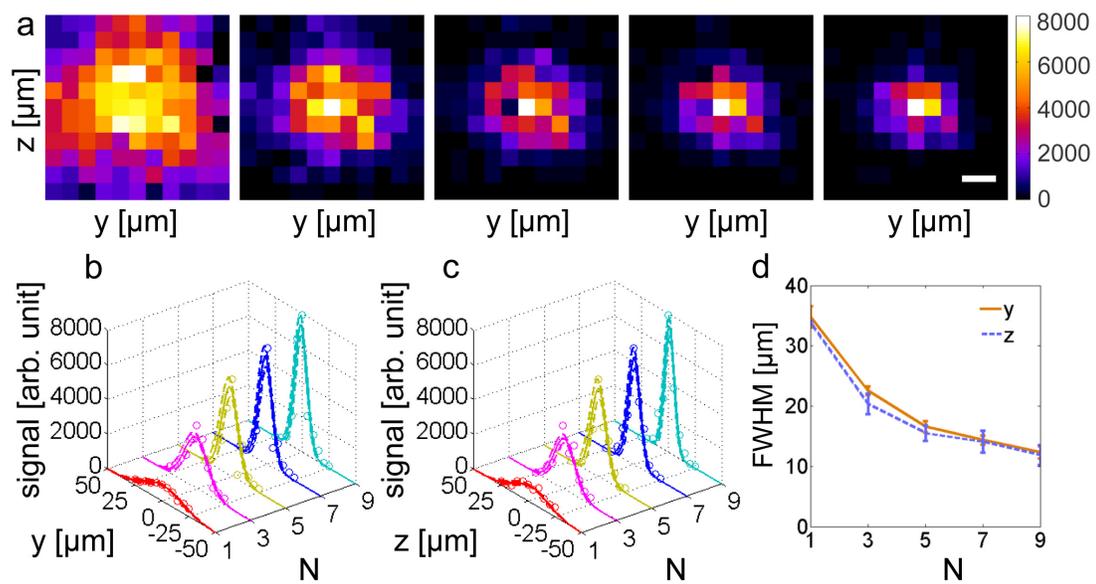

Supplementary Figure 2

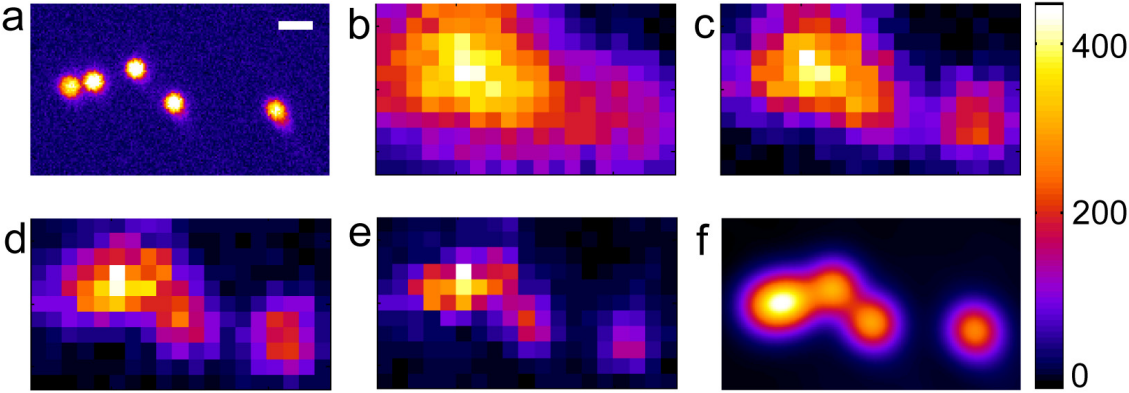